\documentclass{article} 
\usepackage{nips13submit_e,times}
\usepackage{hyperref}
\usepackage{url}
\usepackage{graphicx} 
\usepackage{subfigure} 
\usepackage{amsmath}
\usepackage{amsfonts}
\usepackage{amsopn}
\usepackage{ifthen}
\usepackage[margin=1in]{geometry}

\newcommand{\xb}[0]{\mathbf{x}}
\DeclareMathOperator*{\argmin}{\arg\!\min}

\title{Revisiting Large Scale Distributed Machine Learning}

\author{
Radu C. Ionescu\\
\hspace{0.5cm} School of Computer and Communication Sciences\\
radu-cristian.ionescu@epfl.ch
}

\nipsfinalcopy 

\begin{document}

\maketitle

\begin{abstract}
Nowadays, with the widespread of smartphones and other portable gadgets equipped with a variety of sensors, data is ubiquitous available and the focus of machine learning has shifted from being able to infer from small training samples to dealing with large scale high-dimensional data. In domains such as personal healthcare applications, which motivates this survey, distributed machine learning is a promising line of research, both for scaling up learning algorithms, but mostly for dealing with data which is inherently produced at different locations. This report offers a thorough overview of and state-of-the-art algorithms for distributed machine learning, for both supervised and unsupervised learning, ranging from simple linear logistic regression to graphical models and clustering. We propose future directions for most categories, specific to the potential personal healthcare applications. With this in mind, the report focuses on how security and low communication overhead can be assured in the specific case of a strictly client-server architectural model. As particular directions we provides an exhaustive presentation of an empirical clustering algorithm, k-windows, and proposed an asynchronous distributed machine learning algorithm that would scale well and also would be computationally cheap and easy to implement. 

\end{abstract}

\section{Introduction}

Machine learning offers a wide variety of algorithms for mining and prediction and traditionally the bottleneck was the limited amount of data available, with larger datasets being necessary for high accuracy \cite{megainduction}. However, nowadays the limiting factor is the problem of learning algorithms computational time not scaling well with the data size, making learning from large datasets in a reasonable time impossible. In this context, distributed learning is a promising line of research since allocating the learning process among several computing nodes is a natural way of scaling up learning algorithms, allowing us to deal with both memory and time constrains. Moreover, it allows to deal with data sets that were naturally produced from many physically distributed locations  and/or where transmitting large amounts of data to a central processing unit is expensive or impractical, situations which is becoming more frequent in many real applications \cite{peteiro2013survey}. In some cases the data might contain sensitive information, such as medical or financial records, or data that by nature can not be exchanged with the outside of a company, but still companies would like to exchange knowledge or cooperate, for example, in preventing fraudulent intrusion \cite{kargupta2000collective}. At a high level, distributed learning is the cooperation of experts, each of which has a different perspective based on own observed data,  and in which by message passing the agents communicate to improve the quality of the learned knowledge. Working with mobile agents adds to the challenges of coordination and privacy another stringent factor, low communication overhead.

The potential of machine learning to healthcare application has been recognized from some time \cite{kaur2006empirical, stuntebeck2008healthsense} and also the benefit of multi-agent systems \cite{nealon2003agent}, but in recent years the diversity of sources and the amount of available information, more than 420 million radiological images are generated in US alone \cite{al2015efficient}, have created the need to adopt distributed machine learning techniques \cite{kashyap2015big, slavakis2014modeling}. Furthermore, with the rise of health-tracking apps for mobile phones,  user activity and sensor data are becoming widely available, but the data is mostly used to generate simple statistics and visual attractive activity history plots. Even if efforts are made to incorporate machine learning methods that use behavioral data  to generate personalized feedback \cite{rabbi2015automated},  applications  that work by combining users information to increase accuracy of machine learning methods and that can infer and suggest actions beyond user input, such as in collaborative filtering, are just now emerging, e.g. \cite{jiang2014intelligent}. This subject is just beginning to attract more attention with the creation of Pittsburgh Health Data Alliance \cite{pitshealth} and MRC Health eResearch Centre \cite{mrchealth} and many more focused research centers \cite{jensen2012mining}. Potential to extract useful public-health data goes beyond electronic health records (EHR), with results emerging from web search queries \cite{ginsberg2009detecting} and even social media \cite{dredze2012social}. With the emergence of cloud applications for health data \cite{microsofthealth}, privacy of sensitive health data is becoming more of a requirement \cite{carrion2012personal,verykios2004state}. In this paper we will revisit some of the key advances in distributed machine learning in the hope that they will prove valuable for the research performed in personal healthcare applications.

In the rest of this report we will provide the general problem we are trying to solve in Section 2 and we will present the distributed machine learning algorithms classified based on if a label for the data points is provided or not in either supervised or unsupervised  in Section 3, respectively in Section 4. For supervised learning we will look at the parametric methods of linear regression and Support Vector Machines (SVM), non parametric methods of Gaussian Processes (GP) and probabilistic graphical models (PGM). For unsupervised learning we will focus just on basic clustering methods. Before concluding we present in Section 5 a general implementation for large scale data and our approach to an asynchronous distributed machine learning algorithm.

\section{Preliminaries}
\label{sec:preliminaries}

In this section we will introduce the basic model we will be focusing in the rest of the report. In classical machine learning we are provided with a dataset $\mathcal{D}$ consisting of $N$ points $\xb_i \in$ $\mathbb{R}^{n}$ and in the case of supervised learning we also receive a real value or class label for each point $y_i$ (or we will need to infer it in the case of unsupervised learning). The goal is to find a set of parameters $\theta$ that will minimize a loss function $f(\xb,y \vert \theta)$ over the entire dataset $\theta^{\star} = \underset{\theta}{\argmin} \sum_{i=1}^{N} f(\xb_i,y_i \vert \theta )$. In distributed machine learning we will consider we have $K$ nodes (or agents), each provided with a dataset $\mathcal{D}_k$ consisting of $N_k$ points $\xb^{(k)}_i \in$ $\mathbb{R}^{n}$ and respective labels $y^{(k)}_i$ if provided and the goal is for each agent to learn the same parameter $\theta^{(d)\star} = \underset{\theta}{\argmin} \sum_{k=1}^{K} \sum_{i=1}^{N_k} f(\xb^{(k)}_i,y^{(k)}_i \vert \theta )$. There might be cases were we would want instead of a global optimizer, to have each node learn local optimizers specific to the dataset they have or even a mixture of local and global optimizers, but this will not be the focus of this article. In the case were $\sum_{k=1}^K N_k = N$ and $\cup_{k=1}^{K} \mathcal{D}_k = \mathcal{D}$, the distributed algorithm ideally should converge to the same set of parameters, i.e. $\theta^{(d)\star} = \theta^{\star}$. For evaluation of the performance of the algorithms, in the case the last condition is satisfied, we have the same performance of the non-distributed learning methods. If this is not the case then we need to define a measure of how far the distributed algorithms result diverges from the true one, considered here the non-distributed algorithm. Because we can not have the pie and eat it to, we will incur some overhead from performing the computations distributed, in which case we need to compare performance in complexity, computation time, memory or communication extra costs. Privacy will be also a concern if transmitting $(\xb_i,y_i)$ is forbidden.

\begin{figure}[!htbp]
\centering
   \subfigure[  Sensor network]{\includegraphics[width=5cm]{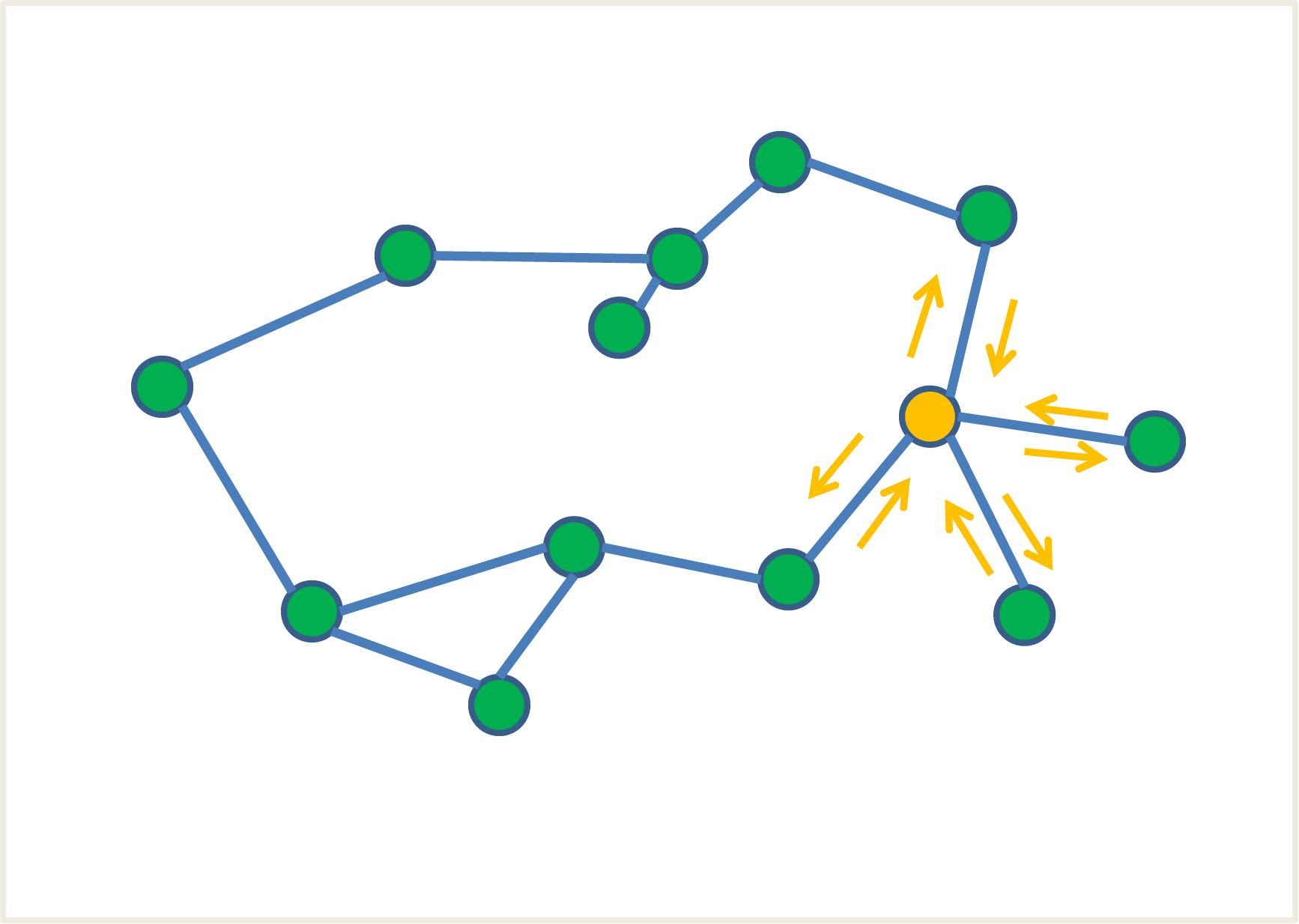} \label{fig:sensor}}
   \subfigure[  Hierarhical network]{\includegraphics[width=5cm]{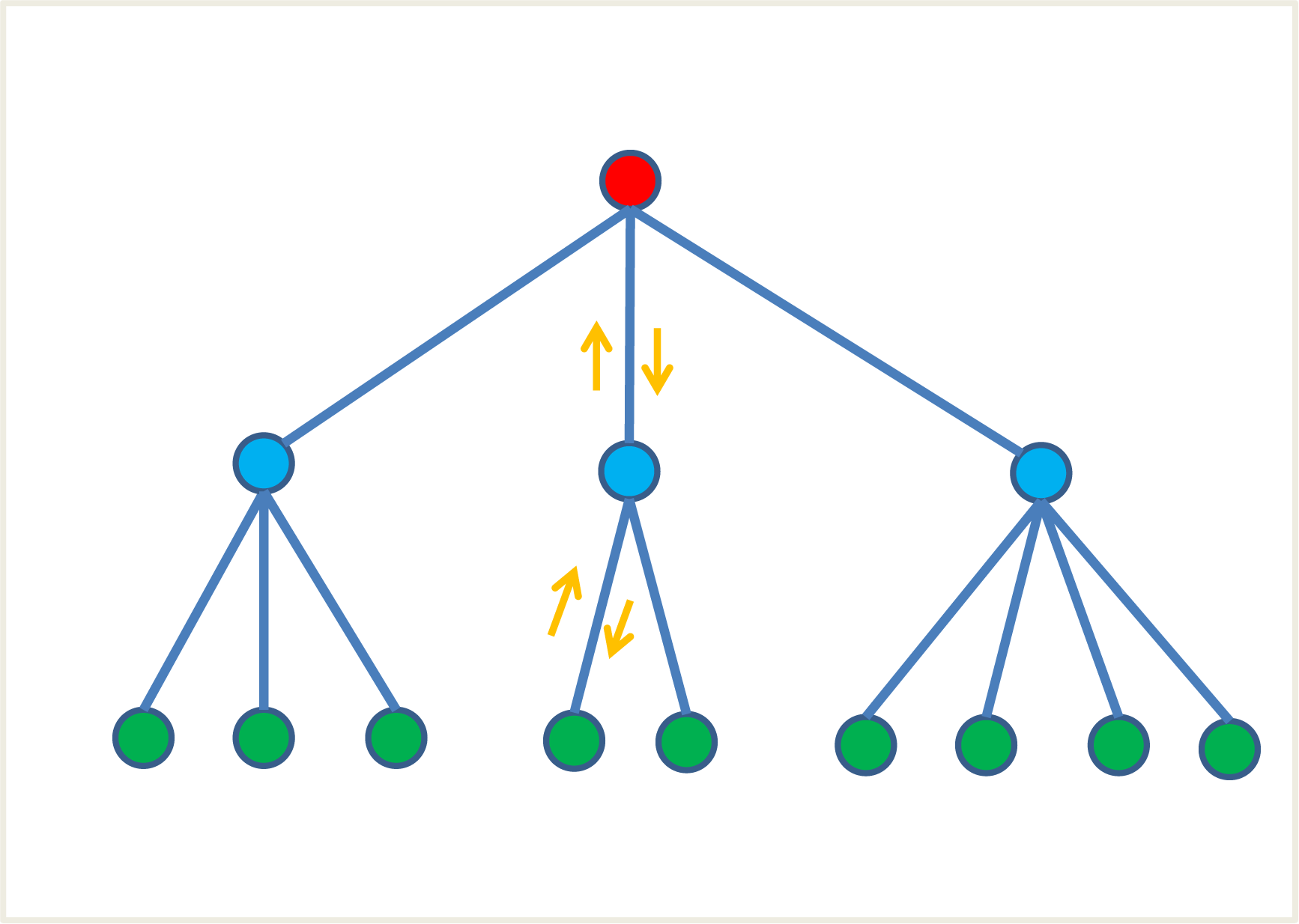} \label{fig:hierarchical}}
   \subfigure[  Central server network]{\includegraphics[width=5cm]{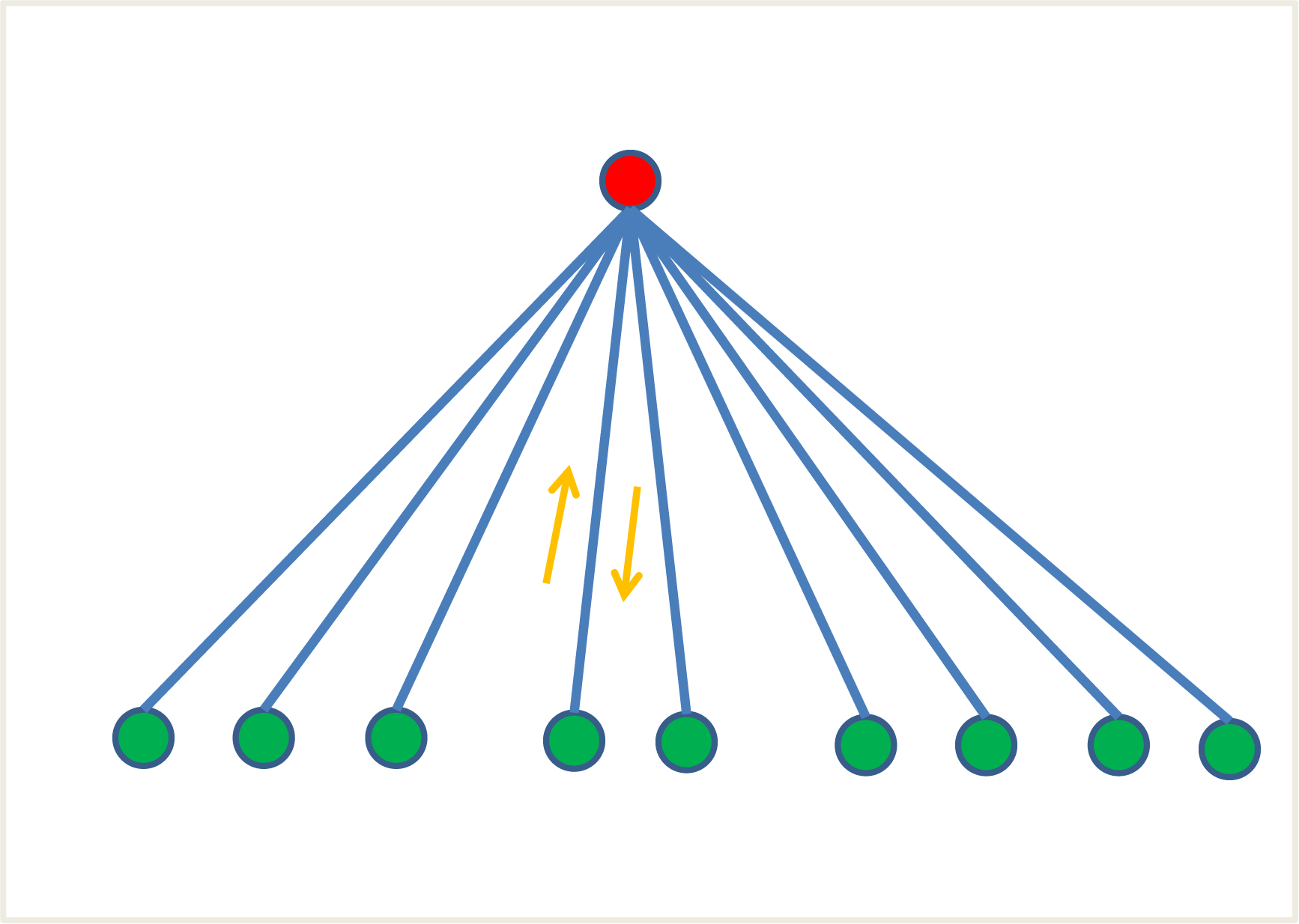} \label{fig:central}}
   \caption{Possible types of network topologies, where the green circles denote data nodes and the yellow arrows show example of information network flow \label{fig:network}}
\end{figure}

From the point of communication between nodes, the network  architecture can be categorized in three scenarios, depicted in Fig.~\ref{fig:network}. In the case of sensor networks \cite{predd2005distributed} a node is allowed to communicate only with its immediate logical or physical neighbors, whereas in the of hierarchical networks the connectivity graph resembles a tree structure and nodes are allowed to communicate only with its immediate ancestor and children. Usually in this case the leaves are the data nodes where the distributed dataset $\mathcal{D}_k$ resides and upper level nodes are used for aggregation and collaboration. The central server network is a special case of hierarchical architecture in which we have a two level deep tree. The focus of this report is to look at distributed machine learning algorithms in the scenario \ref{fig:central}, although a hardware setting like in \ref{fig:central} can have a logical structure like in \ref{fig:hierarchical}. It is not straightforward how an algorithm designed for \ref{fig:sensor} can be adapted to \ref{fig:central}, but a possible starting point is to create a common neighbor designated as the central processing unit.

\section{Supervised learning}
\label{sec:supervised}

\subsection{Linear regression}

For linear regression we have $y_i \in \mathbb{R}$ and $ f(\xb_i,y_i \vert \theta ) = 0.5 \Vert y_i - \mathbf{a} \xb_i - b\Vert_2^2$, where $\theta$ is $\mathbf{a} \in \mathbb{R}^{n}$ and $b \in \mathbb{R}$.

\cite{parikh2014block} and \cite{agarwal2014reliable} provide parallel solutions to logistic regression (with regularization) that hinge upon an \textit{Allreduce} function (from MPI architecture). This function can be simulated by a two step communication with a central server, first each node sends to the server the current local estimate $\theta^{(k)}$ and then al of the nodes receive back from the server the optimal global parameter $\theta$ as a function of this local estimates. Using a regularization function $g(\theta)$ (if $g(\theta)= \Vert \theta \Vert_1$, the problem is known as LASSO) and introducing the variables $z_i = \mathbf{a} \xb_i + b$, we end up with a constrained convex optimization problem of the form $\theta^{\star},\mathbf{z}^{\star} = \underset{\theta,\mathbf{z}}{\argmin} \sum_{i=1}^{N} f(z_i,y_i ) + \sum_{j=1}^{n+1} g(\theta_j)$ such that $z_i = \mathbf{a} \xb_i + b$ for $1\leq i \leq N$. In the distributed case we introduce additional constrains of the form $\theta^{(k)} = \theta$ and $z_i = \sum_{i=1}^{K} z_i^{(k)}$ to account for nodes having access only to local copies of the variables. Application of the Douglas–Rachford splitting (also known as ADMM) to this optimization problem leads to a three stage algorithm with several proximity functions carried in parallel at each node and two \textit{Allreduce} functions, one for coordination on $\theta$ and one for coordination on $\mathbf{z}$, repeated until convergence. \cite{agarwal2014reliable} proposes use of a quasi-Newton method which reduces the number of iterations, but increases communication cost per iteration. L-BFGS is implemented distributed using just one \textit{Allreduce} function to compute the optimal step and to synchronize state, such that each node maintains the rank-1 updates and L-BFGS is only run locally. Furthermore,local  stochastic gradient evaluations are used. The authors in \cite{peng2013parallel} use a similar approach of operator splitting applied to a Coordinate Descent algorithm. They argue that a greedy selection of update coordinates is better suited for sparse solutions instead of cycling in Gauss-Seidel fashion or random selection and imply it outperforms the parallel version of ADMM \cite{ricecomp}.

Privacy of user data is assured in \cite{ahmadi2010privacy} by sending the empirical  second order statistics. If the user data matrix, where each row is a data point or outcome, are $X$ and $Y$ we can find the global optimizer by computing $\theta = (X^T X)^{-1} X^T Y = W^{-1} V$, where $W=\sum_{k=1}^{K} W^{(k)T} W^{(k)}$ and $V = \sum{k=1}^{K} W^{(k)} Y^{(k)}$, with $W^{(k)}=X^{(k)T} X^{(k)}$ and $Y^{(k)}=X^{(k)T} Y^{(k)}$ where $X^{(k)}$ and $Y^{(k)}$ denote each node partition of data. This work is motivated among others by a health application, but privacy can not be assured when user data is highly correlated or when  privacy of some user data distribution is needed and can be used only for masking exact data points values.


In Section 5, we will detail the method in \cite{li2014communication} which uses as example $\ell_1$-regularized logistic regression.

\subsection{Support Vector Machines}

We will look at Support Vector Machines (SVM) for classification, where $y_i \in \{ -1,1\}$ is defined as the class label and $f(\xb_i,y_i \vert \theta ) =[ 1-y_i  \phi_i^T \theta ]_+$, with $[ v ]=\max(0,v)$, $\phi_i = \phi(\xb_i) \in \mathbb{R}^{p}$ for a given basis function $\phi(\cdot)$, and $\theta \in \mathbb{R}^p$. We have also a regularization function given by $g(\theta) = 0.5 \Vert \theta \Vert_2^2$ and regularization parameter $\lambda$, but different penalization functions can be applied leading to other maximal margin classifiers, e.g. $g(\theta) = \Vert \theta \Vert_1$ for Sparse-SVM. SVM is very popular since the dual of the convex optimization problem needed to learn $\theta$, is a quadratic (least-squares) constrained problem, i.e. $\underset{\alpha \in [0,\frac{1}{\lambda}]^N}{\max} \alpha \mathbf{1}^T - \alpha (Y^T \Phi \Phi^T Y) \alpha^T$, that can be solved by a greedy like algorithm (Sequential Minimal Optimization), it is naturally kernelized by $K = \Phi \Phi^T$ (only positive definite kernels work), and the solution $\alpha$ is sparse. Although $\theta$ can be recovered easily, the decision is taken only considering examples that fall at the boundary, called  Support Vectors  (SVs), $\sum_{i=1}^{N} \alpha_i <\phi(x),\phi_i > = \sum_{i, \alpha_i  \text{is SV}} \alpha_i <\phi(x),\phi_i >$, which makes the SVM method extremely fast for classification of a new example.

A first method that only exchanges the SVs is described in \cite{graf2004parallel} as cascade SVM, where SVs from each node are send to the central node and another round of SVM training is performed on the received SVs. The new  SVs are send back to worker nodes where they are added to the training set and the procedure is repeated  recursively until the SVs from one round to the other do not change. A generalization of the cascade SVM was introduced in \cite{lu2008distributed}, where in a directed graph network topology a node combines the SVs received from predecessors with local SVs and pushes the result to its successors. Use of this algorithms using the MapReduce paradigm is presented in \cite{antoniades2014google} and we are also offered a good introduction to the Hadoop architecture.

The Alternating Direction Method of Multipliers (ADMM) has been used in various scenarios where distributed algorithms are posed as a consensus problem and for SVM we can offer as examples \cite{forero2010consensus}, where the consensus problem is formalized on the primal and allows for an easy extension to an online algorithm, and more recently\cite{bai2014consensus}, where the consensus problem is extended also to using a $\ell_1-\ell_2$ regularizer.  While training is performed independent on each node by solving the dual problem for $\alpha^{(k)}$ of the corresponding local examples, in \cite{wang2010scalable}  average consensus between neighbors is used for classification of a new example without partial training classifiers being combined and it is rerun for each presented example. In \cite{yuan2012scalable} a modified dual problem for sparse SVM learning based on removing the inequality constrain and adding an indicator function to the objective is solved again with the use of ADMM. However it is not transformed in a consensus problem and the distributed algorithm simply hinges upon the parallelization of the proximal operator for separable objective functions naturally arising in machine learning problems. 

The Extreme Support Vector Machine (ESVM) classifier \cite{liu2008extreme} can be trained much faster than other nonlinear SVM algorithms since it requires the inversion of small matrices and in \cite{he2011parallel} parallel and incremental algorithms are proposed, but they require shuffeling of the entire set of points, leaving the problem of distributed computation open. 

Not encountered in the literature review, a consensus problem may be formulated for the dual SVM problem as well, in which a node would set some of its local SVs to zero, i.e. $\alpha^{(k)}_i =0$, to satisfy consensus and minimize the objective function along with the overall loss. In this case, however, the Lagrange multipliers parameters choice may play an important role since consensus would be reached by ignoring nodes with more examples and that would eventually contribute to the margin error in a greater proportion than nodes with less training examples in the margin. At least weights proportional to the number of training examples of each node should be used, although other factors would contribute, such as the actual distribution of examples (a node that contains only non essential SVs should ideally have a weight of zero).


\subsection{Gaussian Processes}

In the following two sections we will not keep our notation and minimization of loss function $f(\cdot)$ framework. In Gaussian Process(GP) regression we define a Gaussian process prior over latent function f as $p(\mathbf{f}\vert X) = \mathcal{N}(\mu(X),K(X,X))$ with $K(X,X)_{i,j} = k(\xb_i,\xb_j)$ for some kernel function $k$ and a likelihood over the data values $\mathbf{y}$ as $p(\mathbf{y}\vert \mathbf{f}) = \mathcal{N}(\mathbf{f},\sigma I_N)$ which corresponds to measurements of the form $\mathbf{y} = \mathbf{f} + \epsilon$ with $\epsilon \sim \mathcal{N}(0,\sigma I_N)$. Prediction $f^{\star}$ at a new point $\xb^{\star}$ is obtained using the fact that $f^{\star}$ and $\mathbf{y}$ are jointly Gaussian 
\begin{equation*}
\begin{array}{c c c}
\left[ \begin{array}{c} \mathbf{y} \\  f^{\star}\end{array}\right] & \sim & \mathcal{N}\left( \mu\left( \left[ \begin{array}{c} \mathbf{y} \\  f^{\star}\end{array}\right] \right)\; , \; \left[ \begin{array}{c c} K(X,X)+\sigma I_N  & K(X,\xb^{\star}) \\    K(\xb^{\star},X) & K(\xb^{\star},\xb^{\star}) \end{array}\right] \right)
\end{array}.
\end{equation*}
The marginal $p(f^{\star}\vert \mathbf{y})$ is a Gaussian with mean $\mu^{\star}$ and variance $\sigma^{\star}$ given by
\begin{align*}
\mu^{\star} &= \mu(\xb^{\star}) + K(\xb^{\star},X)  \left[ K(X,X)+\sigma I_N\right]^{-1} \big(\mathbf{y} - \mu(X)\big) \\
\sigma^{\star}&=K(\xb^{\star},\xb^{\star})-K(\xb^{\star},) \left[ K(X,X)+\sigma I_N\right]^{-1} K(X,\xb^{\star})
\end{align*}
GP is a non-parametric method, however we have a set of hyperparameters $\theta$ that define the process, e.g. the kernel might depend on some scale variable. To estimate the hyperparametes we maximize the marginal log-likelihood $p(\mathbf{y}\vert X,\theta) = -0.5\big(\mathbf{y} - \mu(X)\big)^T \left[ K(X,X)+\sigma I_N\right]^{-1} \big(\mathbf{y} - \mu(X)\big)^T-0.5 \log\det( K(X,X)+\sigma I_N)$ \cite{williams2006gaussian}

A practical limitation of Gaussian Processes(GP) is that training and predicting have $\mathcal{O}(N^3)$, respectively $\mathcal{O}(N^2)$ running times.In sparse GP, the true posterior GP can be approximated by averaging  over a small set $q$ of auxiliary inducing variables $\mathbf{f}_q$ drawn from the same GP prior as the training function values  evaluated at pseudo-inputs $X_q$ independent from the training inputs and that ideally should be sufficient statistics for $\mathbf{f}$ ($y^{\star}$ and $\mathbf{f}$ are independent given $\mathbf{f}_q$) and thus $\mu^{\star}$ and $\sigma^{\star}$ are now computed using $X_q$, instead of $X$.  \cite{titsias2009variational} selects the inducing points by minimizing the KL divergence between actual and computed joint distribution of $\mathbf{f},\mathbf{f}_q$.  Recently \cite{gal2014distributed} used the sparse approximation GP introduced in \cite{titsias2009variational} to carry the computation in an embarrassing parallel model on each node and to coordinate uses a central node to accumulate partial results and redistribute back the aggregated value to the network of nodes. 

Computations can also be distributed by using independent local models, where predictions are made by weighting the predictions of local experts. One approach called Mixture-of-experts (MoE) assumes there are $P$ independent experts  $\mathbf{f}^{(p)}\sim\mathcal{N}(0,k^{(p)}(X_p,X_p)$, each with a set of hyperparameters and an auxiliary latent variable $\mathbf{z}_n$ associated which each datapoint that identifies the expert the noisy corrupted measurement $y_n = f_{z_n}(\xb_n)+\epsilon_n$. comes from. We can have hard assignments where one datapoint comes form a single expert or soft assignments where $\mathbf{z}_n$ indicates the proportion of one expert in the mixture of each datapoint. MoE has the drawback it requires  Markov Chain Monte Carlo (MCMC) or variational approximations to assign data points to GP experts and, in order to bypass this problem, \cite{nguyen2014fast} proposed a method  in which each expert is augmented with a set of inducing points and assignment is done by proximity to experts with the use of a MAP estimator. The estimator boils down to $\hat{z}_n = \underset{p \in [1,\dots,P]}{\argmin}(\xb_n - \mathbf{m}_p)^TV^{-1}(\xb_n - \mathbf{m}_p)$, where $\mathbf{m}_p$ and $V$ (diagonal) are quantities based on each expert set of inducing points.

The Product-of-GP (PoE) models multiplies independent GP experts contribution (naturally weighting each contribution), but tend to be overconfident \cite{ng2014hierarchical}. The marginal likelihood in PoE can be factorized according to the local datasets (but not necessarily) as $p(\mathbf{y}\vert X,\theta) \approx \prod_{k=1}^{K} p(\mathbf{y}^{(k)}\vert X^{(k)},\theta)$, which transforms the objective function used for training (log-likelihood maximization) in $K$ separable. After training the prediction can be computed as $p(f^{\star}\vert \xb^{\star},Y) = \prod_{k=1}^{K} p(f^{\star}\vert \xb^{\star},Y^{(k)})$  with the solution $\left( \sigma^{\star}\right)^{-2} = \sum_{k} \left( \sigma^{(k)}(\xb^{\star})\right)^{-2}$ and $\mu^{\star} = \left( \sigma^{\star}\right)^{2}\sum_{k}  \left( \sigma^{(k)}(\xb^{\star})\right)^{-2} \mu^{(k)}(\xb^{\star})$. 
The generalized PoE-GP model   \cite{cao2014generalized} models the importance of GP-experts by weighting contribution as $p(f^{\star}\vert \xb^{\star},Y) = \prod_{k=1}^{K} p^{\beta_k}(f^{\star}\vert \xb^{\star},Y^{(k)})$, which reflects in the solution as  $\left( \sigma^{\star}\right)^{-2} = \sum_{k} \beta_k \left( \sigma^{(k)}(\xb^{\star})\right)^{-2}$ and $\mu^{\star} = \left( \sigma^{\star}\right)^{2}\sum_{k} \beta_k \left( \sigma^{(k)}(\xb^{\star})\right)^{-2} \mu^{(k)}(\xb^{\star})$. For $\sum_{k=1}^{K} \beta_k=1$, the model falls backs to the prior outside the range of the data, but the drawback is that in the range of the data it overestimates and with an increasing number of experts it becomes to conservative. However in a central server model coordination to ensure $\sum_{k=1}^{K} \beta_k=1$ is easy to accomplish. The Bayesian Committee Machine(BCM) \cite{tresp2000bayesian} explicty incorporates the GP prior into predictions as $p(f^{\star}\vert \xb^{\star},Y) = \prod_{k=1}^{K} \frac{p(f^{\star}\vert \xb^{\star},Y^{(k)})}{p^{M-1}(f^{\star}\vert \xb^{\star})}$, which reflects in the solution as  $\left( \sigma^{\star}\right)^{-2} = \sum_{k} \left( \sigma^{(k)}(\xb^{\star})\right)^{-2} + (1-M)\sigma_0^{-2}$ where $\sigma_0$ is the prior variance of $p(f^{\star})$. 
The advantages of this model is again that the correction term ensures a consistent model that falls back to the prior, but it may become inconsistent if a GP expert relies on a small number of datapoints. In \cite{deisenroth2015distributed} proposed the generalized Bayesian Committee Machine (gBCM), which combines the generalized PoE-GP model use of additional weights to model importance of experts with the BCM model  which incorporates the GP prior when combining predictions. The marginal likelihood becomes $p(f^{\star}\vert \xb^{\star},Y) = \prod_{k=1}^{K} \frac{p^{\beta_k}(f^{\star}\vert \xb^{\star},Y^{(k)})}{p^{-1+\sum_{k}\beta_k}(f^{\star}\vert \xb^{\star})}$, which reflects in the solution as  $\left( \sigma^{\star}\right)^{-2} = \sum_{k} \beta_k \left( \sigma^{(k)}(\xb^{\star})\right)^{-2} + (1-\sum_{k}\beta_k)\sigma_0^{-2}$ where $\sigma_0$ is the prior variance of $p(f^{\star})$ .The advantage of the gBCM is that it avoids the drawbacks of the BCM which suffers from weak experts and that of the PoE-GP that overestimates the variance and also can be layered in an arbitrary depth tree structure with each leaf representing a GP expert.

Rather than forcing the nodes to collaborate in building a common GP model (requires central node of coordination in above mentioned sources),one can imagine an approach in which each node builds a local MoE model using own data (one GP expert) and information shared by neighbors/rest of network (other experts).

\subsection{Probabilistic graphical models}

The most common types of probabilistic graphical models are Bayesian Networks (BNs) and Markov Random Fields (MRFs). The BNs can be visualized as a directed acyclic graph where the nodes are given by random variables and edges correspond to influences of one  Random Variable (RV) over another and each RV is associated with a conditional probability distribution given its parents. MRFs can be visualized in the same way only as a undirected graph. As in the case of GPs, graphical models provide an approach to deal with uncertainty and have become an extremely popular tool for modeling uncertainty. 

The computation of the gradient for the Maximum Likelihood Estimator (MLE) in undirected probabilistic graphical models is difficult since while the data term is easily computed, the model term requires evaluating a sum with exponentially many terms over the model distribution. MCMC is a common approach \cite{andrieu2003introduction}, but it is hard to implement in a distributed fashion. Replacing the MLE with a composite estimator, the Maximum Pseudo-Likelihood Estimator (MPLE), in which the conditional distribution of a RV depends only on the neighbors in the Markov Graph, then the gradient becomes data-dependent only, but the same parameter needs to be shared across multiple factors (not distributed friendly computation). In \cite{liu2012distributed} treat it as a consensus optimization problem and solve it using ADMM in a general setting. In \cite{mizrahi2014linear} the composite estimator is created by defining an MRF for a all the variables in a clique q as well as all the variables sharing one neighbor and creating thus a sub-problem for each maximal clique,consensus being achieved in one step, but requires a restrictive factorization. Further in \cite{mizrahi2014distributed}, they continue and are able to import the general conditions from \cite{liu2012distributed} and prove the MPLE as a special case.

\section{Unsupervised learning }
\label{sec:unsupervised}

\subsection{Distributed clustering}

We will now return to the framework of using a loss function $f(\cdot)$, with the particularity that the class labels $y_i$ are not provided and need to be inferred. The purpose of clustering is to find $D$ clusters, each with a prototype vector $\mu_d$, and assign the data points $\xb_n$ to this cluster such that a distance measure between the prototype vectors and the datapoints assigned to them is globally minimized. In the case of hard assignments $f(\xb_n) = \sum_{d=1}^{D} r_{nd}\Vert \xb_n - \mu_d\Vert_2^2$ and the assignment to cluster $d$ is given by $r_{nd}=1$ for the binary variable $r_{nd}$. The optimization algorithm is known as Expectation-Maximization (EM) and consists of iteratively applying two successive stages: in the E-step $\mu_d$ is kept constant and the optimization is carried on $r_{nk}$ which corresponds to assigning datapoins to clusters, in the M-step the assignments $r_{nk}$ are fixed and the optimization is carried on the cluster centroids $\mu_d$. The algorithm is guaranteed to converge to a local minimum. After the training is performed a \textit{K-d tree} can be constructed to minimize the time it takes a newly point to be labeled as belonging to a particular cluster. This is a binary tree where the branches are separated by $c_i<h$ and $c_i\geq h$, for some $i$ dimension and a value $h$. The leaves contain information about which Voroni regions intersects the subspace defined by each of the leaves. This tree can be constructed using statistical information or by using just the training sample data \cite{ramasubramanian1992fast}. Since ultimately new datapoints could influence the structure of the clusters, in \cite{telgarsky2010hartigan} analyzes the Hartigan method (which resembles a submodularity argument)  comparing it to the k-means online version, but leaves a lot of open questions regarding performance optimization and potential interesting results.

For the case of soft assignments a more probabilistic approach has to be followed. One approach is to use a Gaussian mixture model (although any other conjugate distributions pairs could be used), where the marginal distribution $p(\xb_n) = \sum_{d=1}^{D} \pi_{d} \mathcal{N}(\xb_n|\mu_d,\Sigma_d)$ and a latent variables $\mathbf{z}_n$ with the marginal (or prior) $p(z_{nd}=1)=\pi_d$ is used. The cluster soft assignment is given by the conditional probability $p(z_{nk}=1\vert  \xb_n)$ known as responsibilities. Training is performed by maximizing the log-likelihood $\ln p(X|\pi,\mu,\Sigma) = \sum_{i=1}^{N} \ln \left(  \sum_{d=1}^{D} \pi_{d} \mathcal{N}(\xb_i|\mu_d,\Sigma_d) \right)$ with the use of the EM algorithm by optimizing the responsibilities in the E-step and Gaussian experts mean and variance, and prior of the latent variables $\mathbf{z}_n$ in the M-step.

Considering homogenous data distributed across nodes with each node dataset having the same number of clusters and clusters across nodes having the same underlying distribution, \cite{lazarevic2000distributed} construct the convex hulls of the clusters and then averages them at a central node. Considering heterogeneous data, \cite{januzaj2004dbdc} summarizes data in a node by a set of points, such that every point has a minimum number of neighbors and all neighborhoods do not overlap, referred as DBSCAN \cite{ester1996density}. Such a summarization has  low transmission cost, as the number of transmitted representatives to a central server for global aggregation is much smaller than the cardinality of the complete local datasets. Using this representatives, the global clustering is run at the central server using a similar non-distributed algorithm, adapted from DBSCAN. Since the model is the same in each agent when dealing with homogenous data, ADMM can also be used in a consensus problem for the k-means centroids \cite{forero2008consensus}.

In \cite{hajiee2010new} the problem of transmission cost is resolved by dynamically constructing the clusters locally based on a parameter $T$ which defines the maximum radius of a cluster and forming new clusters for points outside, then iterating in transmitting the centroids with summary statistics to a local server which decides if merging of clusters is necessary and sending back aggregated summaries. 

\subsection{K-windows}

It was suggested to analyze in depth the K-windows algorithm proposed in \cite{vrahatis2002new}. The algorithm consists of constructing an initial window(actually an orthogonal range space or box) for each cluster and then iterating in the same manner as the K-means algorithm, assigning datapoints to clusters in the E-step and centering in the M-step. The algorithm is then continued in a second phase where the clusters are enlarged to incorporate as many as possible suitable examples. The algorithm is highly empirically constructed and although several papers have been published no convergence guarantees or proof of optimality is presented. Many of core components or parameter selection are not precisely explained as for example, from \cite{vrahatis2002new} to \cite{tasoulis2005unsupervised} the enlargement procedure has been changed from maintaining the cluster centroid unchanged within a predetermined region to the number of newly added examples to be above a given threshold. The algorithm also has a third phase added in a later publication in which clusters are combined if the number of overlapping examples is greater than a given threshold or if simply if the windows overlap more than a relative volume. The authors propose that the algorithm can be started with a large number of clusters and using this aggregation procedure it will converge close to an optimal number of clusters, but again no guarantees are provided by using this random initialization procedure. The algorithm has been applied to problems from a variety of fields \cite{tasoulis2004unsupervised}\cite{tasoulis2005clustering}\cite{tasoulis2006unsupervised}\cite{pavlidis2006computational}\cite{tasoulis2006unsupervised2}, but our experiments using synthetic data have shown that the algorithm is not very effective in high-dimensional spaces, partially admitted by the authors in \cite{tasoulis2006oriented} where they use PCA for dimensionality reduction, and also it does not perform well when the clusters are not clearly separable. Another problem observed from running the algorithm is that the precision is high (due to the enlargement of windows procedure), but at the same time the recall rate is high considering that there is no stated solution for how assignment is handled for points that belong to two clusters that are not fit to be merged by the third phase. The benefit of the K-windows algorithm is however it can use a datastructure (see \cite{preparata2012computational} or \cite{procopiuc2003bkd} )which allows points belonging to a particular cluster to be easily accessed and speeds up the M-step considerably, although similar algorithms can be used for k-means (and in particular for k-means using $\ell_\infty$ norm)

Because of the way it is currently formulated, the algorithm is not suited for rigorous analytical tools, but we can make the observation that the Voroni regions resemble to Voroni regions constructed by a k-means algorithm using $\ell_\infty$ norm. Furthermore, the particular way the merging is performed seems to indicate that we are checking if the two clusters come from the same uniform distribution (the "overlapping" volume would be proportional to the probability mass of a uniform RV) and even more would be suited for test statistics which are based on "spacing frequencies"
(i.e., the numbers of observations of one sample which fall in between the spacings made by the other sample which in our case coincide with the set of common points)\cite{holst1981asymptotic}. There are a variety of statistical tests which can be used to accept or reject a hypothesis of two samples coming from the same distribution (which is the definition of a cluster), but this goes beyond the scope of this report. The only point we want to make is that using as priors exponential families from the Laplace, Gaussian or "uniform" distribution the use of $\ell_1,\ell_2 \text{ or respectivley} \ell_\infty$ distance measures in the k-means algorithm, they maximize the ML estimator. This prior are given by $p_{\ell_1}(\mathbf{x}) = \frac{1}{C_1} e^{-\Vert\mathbf{x}\Vert_1}$, where $C_1 = \int_{-\infty}^{\infty}\cdots\int_{-\infty}^{\infty}  e^{-\Vert\mathbf{x}\Vert_1} dx_1 \cdots dx_n=2^n$, $p_{\ell_2}(\mathbf{x}) = \frac{1}{C_2} e^{-\Vert\mathbf{x}\Vert_2^2}$, where $C_2 = \int_{-\infty}^{\infty}\cdots\int_{-\infty}^{\infty}  e^{-\Vert\mathbf{x}\Vert_2} dx_1 \cdots dx_n=\sqrt{\pi}^n$ and $p_{\ell_\infty}(\mathbf{x}) = \frac{1}{C_3} e^{-\Vert\mathbf{x}\Vert_\infty}$. where $C_\infty = \int_{-\infty}^{\infty}\cdots\int_{-\infty}^{\infty}  e^{-\Vert\mathbf{x}\Vert_\infty} dx_1 \cdots dx_n=2^n  n!$, and even if the $\ell_\infty$ does not correspond to an exact uniform distribution, for all practical uses it would be almost indistinguishable.

For this reasons we will now look at results for k-means using the $\ell_\infty$ norm. In an early paper on K-means \cite{macqueen1967some}, MacQueen stated that low valued $p$-metric spaces (e.g. $p=1$) will tend to locate the centroids in areas of large probability, while high valued $p$-metric spaces (e.g. $p=\infty$) will tend to have centers such that we have a spherical covering of the space with minimal radius. In \cite{bobrowski1991c} the hard and soft clustering algorithms for $\ell_1$ and $\ell_\infty$ metric spaces are studied, with results showing that the cluster centers are influenced by the initial assignment although the final results tend to match in most cases. The tests were performed on a small dataset and the behavior observed for $\ell_\infty$ remained an open question. In \cite{groenen2007fuzzy}, a slide variation of the metric distance might make the $\ell_\infty$ norm more robust to outliers (but it is not well defined in our opinion).The authors of \cite{bagirov2015nonsmooth} use this distance and proposes a smoothing technique for solving the non-smooth problem of $\underset{x}{\min}\Vert x- a \Vert_\infty$





In the following we will lay the translation of the empirical  k-windows algorithm to a more mathematically sound form, hopefully to make it easier to analyze in the future. Considering for simplicity that the initial volume $a$ is a square with edge length $2r$ (the rectangle case can be obtained by using a weighted $\ell_\infty$ norm, $\ell_\infty^w = \max\{w_1 \vert x_1 \vert,\cdots,w_d \vert x_d \vert\}$ ), the first phase of the algorithm is

$$\underset{c_k,1\leq i \leq K }{\min} \sum_{i=1}^{n} \sum_{k=1}^{K} \mathbf{1}_{\Vert x_i-c_k\Vert_\infty < r} \Vert x_i-c_k \Vert_2^2 $$

or a K-means algorithm where the E-step is skipped and simply replaced with the cluster assignments $u_{i,k} =\mathbf{1}_{\Vert x_i-c_k\Vert_\infty < r} $ and the M-step remaining the same.

For the second phase for each cluster $k$ and each coordinate $d$ we consider $w_d^{k,\text{old}}$ and $w_d^{k,\text{new}}$ to be modified by some enlargement procedure and $w^{k,\text{old}}$  $w^{k,\text{new}}$ to represent the weight of each cluster. The algorithms loops, computing

$$c_k^{\text{new}}=\underset{c_k }{\min} \sum_{i=1}^{n}  \mathbf{1}_{\Vert x_i-c_k\Vert_{\ell_{\infty}^{w^{k,\text{new}}}} < r} \Vert x_i-c_k \Vert_2^2 $$

until $\text{card}  \big( \mathbf{1}_{\Vert x_i-c_k\Vert_{\ell_{\infty}^{w^{k,\text{new}}}} < r} \big)/\text{card}  \big( \mathbf{1}_{\Vert x_i-c_k\Vert_{\ell_{\infty}^{w^{k,\text{old}}}} < r} \big)<\theta$ (or in same versions $\text{dist}(c_k^{\text{new}},c_k^{\text{old}})<\epsilon$).

Similarly the merging of clusters (can start by looking only at clusters for which $\text{dist}(c_i,c_j)<2 \max\{ \Vert w^i r \Vert_\infty,\Vert w^j r \Vert_\infty\}$) is done by evaluating the ratio 
$\text{card}  \big( \mathbf{1}_{\Vert x-c_i\Vert_{\ell_{\infty}^{w^{i}}} < r} \big)/\text{card}  \big( \mathbf{1}_{\Vert x-c_j\Vert_{\ell_{\infty}^{w^{j}}} < r} \big)$.

The algorithm is used with the $\ell_2$ norm, but as pointed in \cite{tasoulis2007generalizing} it can be adapted to other distance measures,e.g. $\ell_\infty$ which would make the comparison with the $\ell_\infty$ k-means more useful.  \cite{tasoulis2004unsupervised} adapts the k-windows algorithm to a "naive" distributed implementation by merging all overlapping windows send by the local nodes at a central server, regardless of the number of examples in the overlapping region of the two windows and thus there is also no need for reiterating. This approach is oversimplified and in the case where the clusters are not clearly separated it was observed that it often leads to merging of neighboring clusters. While it greatly reduces communication costs, it is a direction that would need to be further researched.

\section{Central information server}
\label{sec:central}
\cite{dean2012large} applied asynchronous stochastic gradient descent to a non-convex problem of training a deep neural network and made use of the adaptive learning rate procedure in \cite{duchi2011adaptive} to increase robustness. The data is partitioned across several nodes that create a local model replica, which communicates with a parameter server by pulling current parameters and pushing gradients (computed on a mini-batch). The use of a parameter server is further used in \cite{li2014communication} to create a communication efficient algorithm when using sparse data in high-dimensional space and provide a converge analysis for convex objective functions. 

We propose the following algorithm in which the server in iteration $t$ when a node would push a computed parameter $\theta$ the server would record this as $\theta_{t} \leftarrow \theta$ and would send to the node the parameter $\theta_{t-1}$ from memory. Any machine learning algorithm $F(\cdot)$ chosen to run on each of the nodes would be effectively seen as running in isolation on the local dataset, where the node k after receiving back the parameter from the server would update $\theta^{\text{new}} \leftarrow F^{(k)}( \theta^{\text{received}})$ and on completion would push this result to the central server. The superscript $(k)$ means that local datasets $X_k$ (and $Y_k$) are used by the algorithm.

Let us assume for now that the $K$ nodes would contact the server in a Round Robin fashion and that $\theta_0$ (central server) is initialized to $\theta^{\text{init}}$ and this value would be used also as starting point on the first node contacting the server (node 1). In this case, the second node contacting the server (node 2) would use the parameter $\theta_1$ which was computed using the local dataset of node 1 as a starting point and would update it to reflect also its local dataset. The procedure would continue with every node building on top of the results for its predecessor, ending with an equivalent update of the form $\theta_{t} \leftarrow F^{(t \mod K)}\Big( \dots F^{(2)}\big( F^{(1)}(\theta_0)\big) \dots \Big)$. If $F(\cdot)$ is a first order method based on a convex objective  this is equivalent to a mini-batch gradient descent algorithm. In this case even if the final parameters stored in each of the nodes are not the same, if we guarantee convergences of the non-distributed algorithm, then they are all in a $\epsilon$-neighborhood of the optimal solution. If the datasets themselves are large, $F(\cdot)$ could also be stochastic in nature and the algorithm would converge as long as the non-distributed algorithm converges under a stochastic mini-batch algorithm.  

The extension to an asynchronous version is now straightforward if indeed the learning method is backed by a non-distributed algorithm for which a stochastic mini-batch algorithm exists. Whenever a nodes finishes it computations it contacts the server and ask for the new parameter $\theta$ and in the unlikely case of two nodes contacting the server at exactly the same moment, ties can be broken arbitrarily. As in the synchronous version, every node builds on top of the results for its predecessor, but the equivalent update has the form $\theta_{t} \leftarrow F^{(S_t)}\Big( \dots F^{(S_2)}\big( F^{(S_1)}(\theta_0)\big) \dots \Big)$, where $S_1,S_2,\dots , S_t$ are discrete RV from a distribution $\mathcal{S}$ with the support $\{1,\dots,K\}$. By ensuring that $p(S=i)>0$ for every $1\leq i \leq K$ and $S \sim \mathcal{S}$, which is equivalent to saying that there exists no node that will never contact the server, it can be proven that the algorithm indeed converges. The actual distribution $\mathcal{S}$ is dependent on the local datasets, e.g. number of examples which practically at least influence memory retrieval time.


The analysis is not completed as we can imagine that online learning algorithms can be used and we only wished to detail this method which to the best of our knowledge is new and would resolve a stringent problem in distributed algorithms (not machine learning solely) efficient use of asynchronicity for parallel algorithms. It has the advantage that there is no wasted local computation, but also that it is easy to implement and also to provide convergence guarantees (at least if  we start from non-distributed algorithm for which a stochastic mini-batch algorithm exists).

\section{Discussion}

Although a lot of progress has been made in the past decade on adapting machine learning algorithms to distributed settings, a lot of questions still remain open. For us the most stringent question is if we can easily import the results from the senor network, obtained with great effort during the hype of wireless sensor network, into a central node setting. Most of the distributed algorithms are obtained from trying to parallelize their successful non-distributed counterparts, but for most of the times we have just empirical adaptations, without proof of results equivalence and how the convergence rate is affected by this tradeoff. While communication costs for most supervised learning algorithms is reduced and also various methods of masking actual data points to ensure  privacy exist, simply because most of the times we can get away with just transmitting the set of parameters or hyperparameters, for unsupervised clustering it still constitutes into a great challenge, especially in the case of heterogeneous distributed data.

We have proposed future directions for both supervised and unsupervised learning, ranging from simple linear logistic regression to graphical models and clustering. The report focused on how security and low communication overhead can be assured in the specific case of a strictly client-server architectural model, specific to the potential personal healthcare applications, but at the same time there are other influential factors,e.g. performance and accuracy of a global optimizer versus parameters adapted to local data or a mixture of both.

The empirical clustering algorithm, k-windows, is not as it is now a solution, many of the core components  are not precisely defined or supported by analytical choices and influence of parameter selection is not obvious. From our experiments it does not perform well in high-dimensional spaces and does not perform well when the clusters are not clearly separable, especially in the distributed version where the aggregation procedures leads to merging of neighboring clusters.

On the other hand we proposed an asynchronous distributed machine learning algorithm that would scale well and also would be computationally cheap and easy to implement. If the learning method is backed by a non-distributed algorithm for which a stochastic mini-batch algorithm with convergence guarantees exists, adapting it to our method is straightforward and we can also import all of the  analytical results, including convergence rate which surprisingly would remain the same as in its non-distributed counterpart. This is an amazing result considering the fact most of the times there exists a tradeoff between parallelization and convergence rate.

%
%

\newpage
\bibliography{bib}{}
\bibliographystyle{plain}

\end{document}